# A Survey on Information Retrieval, Text Categorization, and Web Crawling

Youssef Bassil

LACSC – Lebanese Association for Computational Sciences
Registered under No. 957, 2011, Beirut, Lebanon

**Abstract**

*This paper is a survey discussing Information Retrieval concepts, methods, and applications. It goes deep into the document and query modelling involved in IR systems, in addition to pre-processing operations such as removing stop words and searching by synonym techniques. The paper also tackles text categorization along with its application in neural networks and machine learning. Finally, the architecture of web crawlers is to be discussed shedding the light on how internet spiders index web documents and how they allow users to search for items on the web.*

## 1. Introduction

Information Retrieval (IR) is the science of searching for information within relational databases, documents, text, multimedia files, and the World Wide Web [1]. Many users are engaged in the IR field especially reference librarians, professional researchers, political analysts, governmental investigators, and market forecasters. The applications of IR are diverse, they include but not limited to extraction of information from large documents, searching in digital libraries, information filtering, spam filtering, object extraction from images, automatic summarization, document classification and clustering, and web searching.

The idea of searching for information was first mentioned by Vannevar Bush in 1945 [2]. Ten years later the first operational information retrieval systems was introduced. By 1990 several different techniques had been brought in, which process no more than several thousand documents [2]. The breakthrough of the Internet and web search engines have urged scientists and large firms to create very large scale retrieval systems to keep pace with the exponential growth of online data. Figure 1 depicts the architecture of a general IR system. The user first submits a query which is executed over the retrieval system. The latter, consults a database of document collection and returns the matching document.

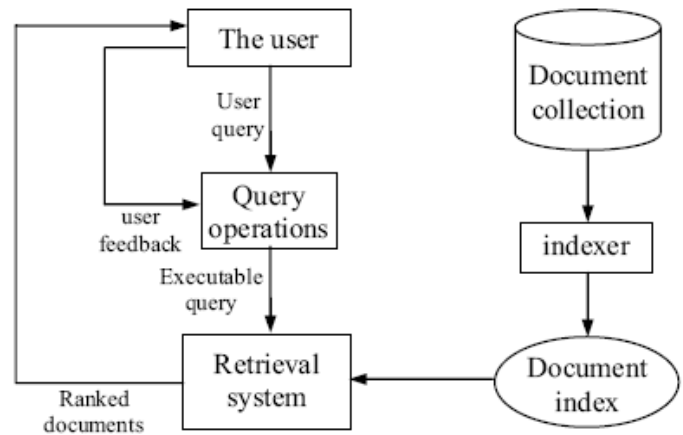

Figure 1 - General IR System Architecture

## 2. Background

Document refers to the unit of text indexed in the system and available for retrieval [3]. A document could be a newspaper, a paragraph of sentences, encyclopaedias entries or a web page. The type of data in a document falls into two categories: structured and unstructured.

Structured Data are data organized and stored in fields, headers and tags, and therefore they are easily usable by a computer; whereas, unstructured data are data with no clear, semantically overt structure, which are harder from their counterparts to be processed and extracted. Almost all data found on computers and on the Internet are unstructured. It may include audio, video, and unstructured text such as the body of an e-mail message, web page, or word processor document.

A collection is a set of documents available for retrieval. A term refers to the lexical item or word that occurs in the document or in the collection. Finally a query represents a user's information need expressed as a set of terms [3].

## 3. Document and Query Modeling

In order to process a document, it should be transformed into a model understandable by an information retrieval system. This section discusses the different method used to model a text document and a user query in an information retrieval system.

### 3.1. The Vector Space Model

The vector space model is a data model for representing documents and queries in an Information Retrieval system. Every document and query is represented by a vector whose dimensions, which are called features, represent the words that occur within them [4]. In that sense, each vector representing a document or query consists of a set of features which denote words and the value of each feature is the frequency or the number of occurrence of that particular word in the document itself. Since an IR usually contains more than one document, vectors are stacked together to form a matrix. Figure 2 shows a single vector populated with frequencies of the words contained in document D.

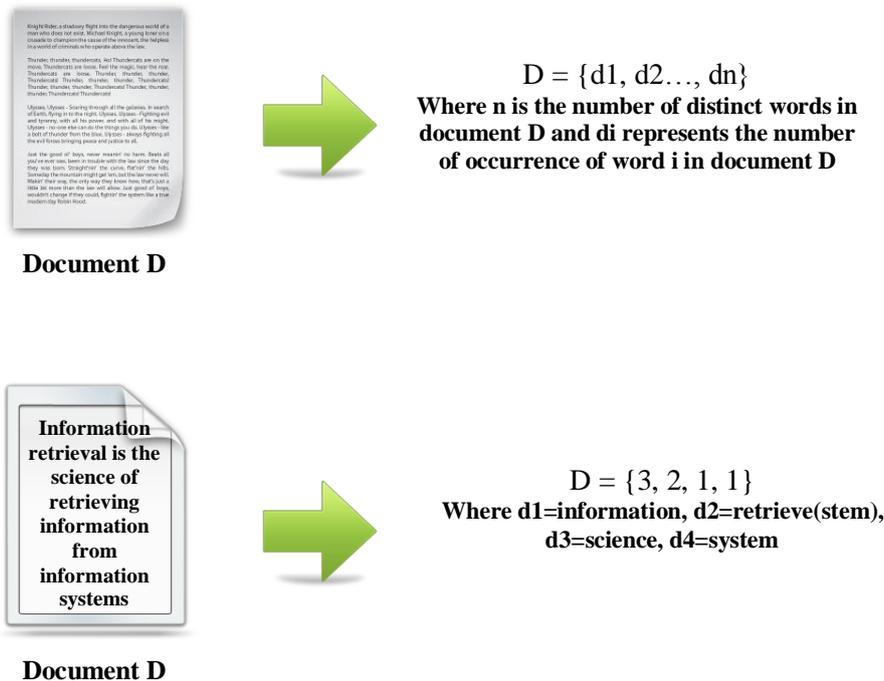

**Document D**

$D = \{d_1, d_2\ldots, d_n\}$
**Where n is the number of distinct words in document D and di represents the number of occurrence of word i in document D**

**Document D**

$D = \{3, 2, 1, 1\}$
**Where d1=information, d2=retrieve(stem), d3=science, d4=system**

Figure 2 – Vector Space Model Example

To make things clearer, let's consider an example of finding the occurrence of words *information*, *processing*, and *language* in three documents *Doc1*, *Doc2*, and *Doc3*

After counting the occurrence of the three words in each of the three documents, *Doc1* is represented by the vector D1(1,2,1), *Doc2* by the vector D2(6,0,1), and *Doc3* by the vector D3(0,5,1)

In order to emphasize on the contribution of the higher value feature, those vectors are to be normalized. By normalization, we simply mean converting all vectors to a standard length. This can be done by dividing each dimension in a vector by the length of that particular vector. The length of a vector can be calculated according to the following equation: *length = sqrt((ax * ax) + (ay * ay) + (az * az))*. Normalizing D1: *length = sqrt((1 * 1) + (2 * 2) + (1 * 1)) = sqrt(6) = 2.449*. Now dividing each dimension by the length: *1/2.449 = 0.41 ; 2/2.449 = 0.81 ; 1/2.449 = 0.41*. Final result would be *D1(0.41, 0.81, 0.41)*. Same applies for D2 and D3 which will eventually result in *D2(0.98, 0, 0.16)* and *D3(0, 0.98, 0.19)*.

Now in order to determine the difference between two documents or if a query matches a document, we must calculate the *cosine* of the angles between the two vectors. When two documents are identical (or when a query completely matches a document) they will receive a cosine of 1; when they are orthogonal (share no common terms) they will receive a cosine of 0.

$$s(\vec{q}_k, \vec{d}_j) = \vec{q}_k \cdot \vec{d}_j = \sum_{i=1}^{N} w_{i,k} \times w_{i,j}$$

Back to the previous example, let's consider a query with corresponding normalized vector Q(0.57, 0.57, 0.57). The first task is to compute the cosines between this vector and our three document vectors.

*Sim(D1,Q) = 0.41*0.57 + 0.81*0.57 + 0.41*0.57 = 0.92*
*Sim(D2,Q) = 0.65*
*Sim(D3,Q) = 0.67*

The previous results show clearly that D1 is the closest to match Q, then comes D3 and then comes D2 (remember that the more the cosine is closer to 1 the more the two vectors are closer)

### 3.2. Term Frequency and Weighting

Obviously, terms or words that have occurred more in a particular document, should have a higher score than the others terms. This score is more specifically called weight and represents how much a given term has an impact and reflects strongly the meaning of the document. The raw frequency of a term within a document is called term frequency [5], and is denoted by $tf_{t,d}$, with the subscripts denoting the term and the document in order.

Raw term frequency as above suffers from a critical problem; all terms are considered equally important when it comes to assessing relevancy on a query. In fact, certain terms have little or no discriminating power in determining relevance. For instance, a collection of documents on the auto industry is likely to have the term auto in almost every document. To this end, a mechanism is introduced for attenuating the effect of terms that occur too often in the collection to be meaningful for relevance determination. An immediate idea is to scale down the term weights of terms with high collection frequency. The idea is to reduce the *tf* weight of a term by a factor that grows with its collection frequency [6]. This proposed idea is called *Inverse Document Frequency* (idf) and is denoted by:

$$idf_j = \log\left(\frac{n}{df_j}\right)$$

$$w_{ij} = tf_{ij} \cdot idf_j$$

Where:

- $idf_j$ is Inverse document frequency.
- $df_j$ is the number of documents in which word $t_j$ occurs in the document collection.
- n is the total number of documents in the document collection.
- $tf_{ij}$ is the term frequency
- $w_{ij}$ is the weight

### 3.3. Extracting and Calculating Term Frequency

We have seen in previous section that the term frequency or the number of occurrence of a word in a document is critical in reflecting the meaning in the document. Hence, we need a way to extract different words from a document and count its frequency. The solution is called *Tokenization*. Tokenization is the process of chopping a document into small units called *tokens* which usually results in a set of atomic words having a useful semantic meaning.

The easiest way to perform tokenization in text document is to split on whitespaces and throwing away punctuations such as commas, full stop, colons, and semicolons. Below is an example of tokenizing a certain input sentence:

```
Input: Friends, Romans, Countrymen, lend me your ears;
Output: Friends | Romans | Countrymen | lend | me | your | ears
```

Conceptually, splitting on white space can also split what should be regarded as a single token [6]. This occurs most commonly with names (San Francisco, Los Angeles) but also with borrowed foreign phrases (au fait) and compounds that are sometimes written as a single word and sometimes space separated (such as white space vs. whitespace). Other cases with internal spaces that we might wish to regard as a single token include phone numbers [(800) 234-2333] and dates (Mar 11, 1983). Splitting tokens on spaces can cause bad retrieval results, for example, if a search for York University mainly returns documents containing New York University.

Other languages make the problem harder in new ways. German writes compound nouns without spaces (e.g., compounds Computerlinguistik – "computational linguistics"; Lebensversicherungsgesellschaftsangestellter – "life insurance company employee"). Such problem can be solved by the use of a compound splitter module, which is usually implemented by seeing if a word can be subdivided into multiple words that appear in a vocabulary.

Another problem in the Chinese language is that the two characters can be treated as one word meaning "monk" or as a sequence of two words meaning "and" and "still." One approach here is to perform word segmentation as prior linguistic processing. Methods of word segmentation vary from having a large vocabulary and taking the longest vocabulary match with some heuristics for unknown words to the use of machine learning sequence models, such as hidden Markov models or conditional random fields, trained over hand-segmented words.

## 4. Document Preprocessing

In order to facilitate the extraction of information from documents, researches have found that preprocessing the document before being exposed to indexing, is so far a good technique which increases the precision of an IR system. In this section different preprocessing technique are tackled and discussed.

### 4.1. Stop Words

Words that are of little value to convey the meaning of the document and which happen to have a high frequency are totally dropped during the tokenization process. These words are called stop words and are generally detected by either their high frequency or by matching them with a dictionary. Below is a stop list of twenty-five semantically nonselective words that are common in Reuters-RCV1.

| a | an | and | are | as | at | be | by | for | from |
| has | he | in | is | it | its | of | on | that | the |
| to | was | were | will | with | | | | | |

Dropping stop words sounds a very good approach to discard some useless redundant words; however this is not the case for some phrases. Imagine a user is searching for "President of the United States" or "Flight to London", the IR system would then search for "President" and "United States" separately or it would search for "Flight" and "London" separately. This would of course return an incorrect result that does not reflect the user's initial query.

### 4.2. Normalization

Token normalization is the process of canonicalizing tokens so that matches occur despite superficial differences in the character sequences of the tokens [6]. The most standard way to normalize is to implicitly create equivalence classes, which are normally named after one member of the set.

### 4.3. Hyphens, Punctuations and Digits

It is sometimes preferable to remove hyphens from the compound words so that co-education would become coeducation. This would eventually map both tokens *co-education* and *coeducation* to the term coeducation, in both the document text and queries, and then searches for one term will retrieve documents that contain either.

Punctuations in the middle of a token are usually discarded so that U.S.A becomes USA after deleting the three periods. For instance this seems to be very reasonable nevertheless if the user searches for a company named C.A.T, he would be upset if the result returns all documents which contain the word *cat*. Digits are usually removed except some specific types, e.g., dates, times, and currencies.

### 4.4. Capitalization and Case-Folding

A typical strategy is to do case-folding by converting all uppercase characters to lowercase characters. This is a form of word normalization in which all words are reduced to a standard form. This would equate between *Door* and door and between *university* and *UNIVERSITY*. This sounds very nice; however the problem arises when a proper noun such as *Black* is equated with the color *black* or it can also equate between the company name VISION and the word vision. To remedy to this, one can only convert to lowercase words at the beginning of a sentence and words located within titles and headings.

### 4.5. Stemming and Lemmatization

Words morphology is exclusively found in every document and query; it comprises inflectional and derivational morphology. *Inflectional* morphology takes as input a word and outputs a form of the same word appropriate to a particular context. For instance talk transformed into talked. On the other hand, *derivational* morphology takes as input a word and outputs a different word that is derived from the input. For instance establish + ment + ary + an + ism [3].

The goal of stemming and lemmatization is to reduce inflection and derivation. Stemming usually chops off the ends of words by the removing derivational affixes. Whereas Lemmatization usually harnesses the use of a vocabulary and morphological analysis of words, in order to reduce the word to its lemma (root of the word). For example taking the word a stemmer would convert all these words *operating operates operation operative operatives operational* to *oper* while a lemmatizer converts them into *operate*. Obviously, a lemmatizer is by far a better approach since it conserves the precision between the query and the document.

Different stemmers exist and some of them can be downloaded from the Internet at no cost. Below is an output produced by four different stemmers:

**Sample text:** *Such an analysis can reveal features that are not easily visible from the variations in the individual genes and can lead to a picture of expression that is more biologically transparent and accessible to interpretation*

**Lovins stemmer:** *such an analys can reve featur that ar not eas vis from th vari in th individu gen and can lead to a pictur of expres that is mor biolog transpar and acces to interpres*

**Porter stemmer:** *such an analysi can reveal featur that ar not easili visibl from the variat in the individu gene and can lead to a pictur of express that is more biolog transpar and access to interpret*

**Paice stemmer:** *such an analys can rev feat that are not easy vis from the vary in the individ gen and can lead to a pict of express that is mor biolog trans and access to interpret*

### 4.6. Synonymy

An IR system should map all words whose meaning is the same to the same word. For instance a query containing the word *car*, the IR system should be able to match this query with all documents containing the word *automobile*. This ultimately increase the rate of the relevant documents returned to the user. To find the synonym of a word, the IR system must consult a thesaurus which is a combination of a dictionary and a collection of correlated terms.

## 5. Evaluation of Information Retrieval

An Information retrieval system is typically evaluated by its ability to return relevant documents and by how much these relevant returned documents are accurate and precise. Retrieving relevant documents is assessed with a recall measure

$$\text{Recall} = \frac{\text{\# of relevant documents returned}}{\text{total \# of relevant documents in the collection}}$$

Determining the accuracy of an IR system is determined by how much the returned documents are exactly relevant. It is assessed by a precision metric

$$\text{Precision} = \frac{\text{\# of relevant documents returned}}{\text{\# of documents returned}}$$

## 6. Text Categorization

Text categorization is the task of classifying a given data instance into a pre-specified set of categories. In other words, given a set of categories, and a collection of text documents, *text categorization* or TC is the process of finding the correct topic for each document [7]. TC has many diverse applications including indexing, web page categorization, spam filtering, and detection of text genre.

Actually two main approaches are in use today for performing text categorization: *knowledge engineering* and *machine learning* (ML).

In *knowledge engineering* a bunch of domain expert people build manually, by hand, rules and categories containing a predefined set a words belonging to a particular domain. This technique is considered labor intensive and requires huge amount of highly skilled of expert knowledge needed to create and maintain the knowledge rules and categories.

In *machine learning* a classifier is built and trained over a set of pre-classified documents whose categories are also predefined. This approach of learning is called *supervised learning* since the predefined documents and categories are set by a human and not deduced automatically by the system. The latter case is called *unsupervised learning* and it is usually employed with text *clustering*.

### 6.1. Document Representation

The common classifiers and learning algorithms cannot directly process the text documents in their original form. Therefore, during a preprocessing step, the documents are converted into a more manageable representation. Typically, the documents are represented by feature vectors. A document is represented as a vector containing a sequence of features and their weights.

The most common *bag-of-words* model simply uses all words in a document as the features, and thus the dimension of the feature space is equal to the number of different words in all of the documents [7]. The methods of giving weights to the features may vary. The simplest is the binary in which the feature weight is either one if the corresponding word is present in the document or zero otherwise. More complex weighting schemes are possible that take into account the frequencies of the word in the document, in the category, and in the whole collection. The most common scheme is the TF-IDF scheme explained previously in this report.

Due to the large number of terms in a document, the number of features is to increase and consequently the dimension of the feature vector. Therefore a good approach is to reduce the number of dimensions and to create a new, much smaller set of features from the original feature set. The preprocessing step that removes the irrelevant words is called *feature selection*. It includes dropping stop words, discarding punctuations, finding synonyms, stemming, and normalizing terms, to create a transformation from the original feature space to another space of much lower dimension.

### 6.2. Knowledge Engineering

In knowledge engineering a domain expert sits and writes manually a set of rules that classify a particular document into a particular category. Furthermore, categories are tabulated and populated with a list of keywords semantically close to the category subject. Despite this approach had shown a great accuracy and precision [7], it remains labor intensive and time wasting since it would take several man-years to acquire knowledge about different domains and to develop and fine tune TC systems.

Below is a procedure of classification rules. Figure 3 shows a collection of categories containing predefined keywords.

*if* (disjunction of conjunctive clauses) formula *then* category *else* ¬category

*If* ((congestion & london) *or*
(windsor & the queen))
*then* UK

*else* ¬China

*If* (feed & chicken)
*then* poultry

*If* ((recount & votes) *or*
(seat & run-off) *or*
(TV ads & campaign))
*then* elections
*else* ¬sports

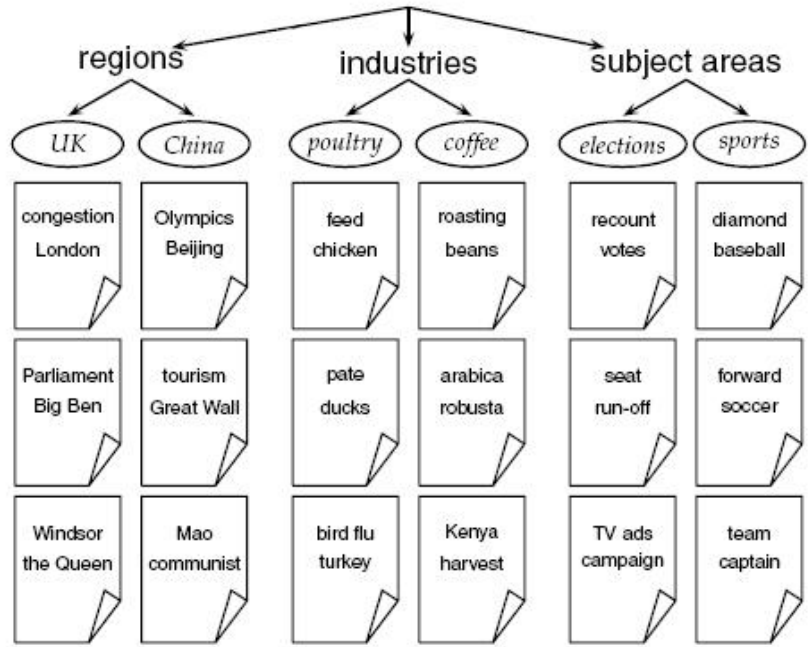

Figure 3 – Categories with Keywords

### 6.3. Machine Learning

In the ML approach, the classifier is built automatically by learning the properties of categories from a set of pre-classified training documents. In the ML terminology, the learning process is an instance of supervised learning because the process is guided by applying the known true category assignment function on the training set.

Four main issues need to be considered when using machine learning techniques to develop an application based on text categorization. First, we need to decide on the categories that will be used to classify the instances. Second, we need to provide a training set for each of the categories. As a rule of thumb, about 30 examples are needed for each category. Third, we need to decide on the features that represent each of the instances. Usually, it is better to generate as many features as possible because most of the algorithms will be able to focus just on the relevant features. Finally, we need to decide on the algorithm to be used for the categorization.

### 6.3.1. Naive Bayes Text Classification

Naive Bayes is a supervised, probabilistic learning method. The Probability of a document $d$ being in class $c$ is computed as:

$$P(d\,|\,c) = \prod_{1 \leq i \leq n_d} P(w_i\,|\,c).$$

Where $P(w_i /c)$ is the conditional probability of term $w_i$ occurring in a document of class $c$. We interpret $P(w_i /c)$ as a measure of how much evidence $w_i$ contributes that $c$ is the correct class. $<w_1, w_2, ..., w_{nd}>$ are the tokens in d that are part of the vocabulary we use for classification and $nd$ is the number of such tokens in $d$.

In text classification, our goal is to find the best class for the document. The maximum a best class in Naive Bayes classification is the most likely or maximum a posteriori and denoted by:

$$c_{\text{map}} = \arg\max_{c \in C} \prod_{1 \le i \le nd} P(w_i | c)$$

### 6.3.2. Support Vector Machines (SVM)

In general, SVM is a linear learning system that builds two-class classifiers.
Let the set of training examples D be $\{(x_1, y_1), (x_2, y_2), ..., (x_n, y_n)\}$, where $x_i = (x_{i1}, x_{i2}, ..., x_{ir})$ is a r-dimensional input vector in a real-valued space, $y_i$ is its class label (output value) and $y_i$ belongs to $\{1, -1\}$. 1 denotes the positive class and -1 denotes the negative class. To build a classifier, SVM finds a linear function of the form:

**f(x) = <w . x> + b**

so that an input vector $x_i$ is assigned to the positive class if $f(x_i) >= 0$, and to the negative class otherwise.

$$y_i = \begin{cases} 1 & \text{if } \langle \mathbf{w} \cdot \mathbf{x}_i \rangle + b \ge 0 \\ -1 & \text{if } \langle \mathbf{w} \cdot \mathbf{x}_i \rangle + b < 0 \end{cases}$$

Hence, f(x) is a real-valued function, $w = (w_1, w_2, ..., w_r)$ is the weight vector. $b$ is called the bias. $<w . x>$ is the dot product of w and x (or Euclidean inner product). Without using vector notation, Equation can be written as:

**$f(x_1, x_2, ..., x_r) = w_1x_1 + w_2x_2 + ... + w_rx_r + b$**

SVM also has some limitations [8]:

1. It works only in real-valued space. For a categorical attribute, we need to convert its categorical values to numeric values. One way to do this is to create an extra binary attribute for each categorical value, and set the attribute value to 1 if the categorical value appears, and 0 otherwise.
2. It allows only two classes, i.e., binary classification. For multiple class classification problems, several strategies can be applied.
3. The hyperplane produced by SVM is hard to understand by users. It is difficult to picture where the hyperplane is in a high-dimensional space.
Thus, SVM is commonly used in applications that do not required human understanding.

### 6.3.3. K-Nearest Neighbor Learning

k-nearest neighbor (kNN) is a lazy learning method in the sense that no model is learned from the training data. Learning only occurs when a test example needs to be classified. The idea of kNN is extremely simple and yet quite effective in many applications. It works as follows: Let D be the training data set. Nothing will be done on the training examples. When a test instance d is presented, the algorithm compares d with every training example in D to compute the similarity or distance between them. The k most similar (closest) examples in D are then selected. This set of examples is called the k nearest neighbors of d. d then takes the most frequent class among the k nearest neighbors. The general kNN algorithm is given as follows:

*Algorithm kNN(D, d, k)*
*1 Compute the distance between d and every example in D;*
*2 Choose the k examples in D that are nearest to d, denote the set by P (∈ D);*
*3 Assign d the class that is the most frequent class in P (or the majority class).*

The key component of a kNN algorithm is the distance/similarity function, which is chosen based on applications and the nature of the data. For relational data, the Euclidean distance is commonly used. For text documents, cosine similarity is a popular choice.

Despite its simplicity, researchers have showed that the classification accuracy of kNN can be quite strong and in many cases as accurate as those elaborated methods. For instance, it is showed in [9] that kNN performs equally well as SVM for some text classification tasks. kNN is also very flexible. It can work with any arbitrarily shaped decision boundaries. kNN is, however, slow at the classification time. Due to the fact that there is no model building, each test instance is compared with every training example at the classification time, which can be quite time consuming especially when the training set D and the test set are large. Another disadvantage is that kNN does not produce an understandable model. It is thus not applicable if an understandable model is required in the application.

### 6.3.4. Neural Networks

Neural network (NN) can be built to perform text categorization. Usually, the input nodes of the network receive the feature values, the output nodes produce the categorization status values, and the link weights represent dependence relations. For classifying a document, its feature weights are loaded into the input nodes; the activation of the nodes is propagated forward through the network, and the final values on output nodes determine the categorization decisions. The neural networks are trained by back propagation, whereby the training documents are loaded into the input nodes. If a misclassification occurs, the error is propagated back through the network, modifying the link weights in order to minimize the error. The simplest kind of a neural network is a perceptron. It has only two layers: the input and the output nodes.

### 6.3.5. Comparison among Classifiers

Given the lack of a reliable way to compare classifiers across researchers, it is possible to draw only very general conclusions in reference to the question which classifier is the best?

According to most researchers, the top performers are SVM, AdaBoost, kNN, and Regression methods. Insufficient statistical evidence has been compiled to determine the best of these methods. Efficiency considerations, implementation complexity, and other application-related issues may assist in selecting from among these classifiers for specific problems. Rocchio and Naive Bayes have the worst performance among the ML classifiers, but both are often used as baseline classifiers.

There are mixed results regarding the neural networks and decision tree classifiers. Some of the experiments have demonstrated rather poor performance, whereas in other experiments they performed nearly as well as SVM.

## 7. Web Crawling and the World Wide Web

The World Wide Web or www for short is a collection of millions even billions of documents written using the HTML language and available online to all Internet users. HTML which is a concatenation for Hypertext Markup Language allows web authors to create web pages that encompass text, images, tables, graphs, links and richer multimedia such as audio, video, and animation. These hypertext pages are served to users through the HTTP (HyperText Transfer Protocol) where they can be displayed using an Internet browser such as Internet Explorer or Firefox.

Web pages are usually interconnected through hyperlinks or simply links. Hyperlinks name and refer to other web pages. Once clicked they navigate the user to another page generally referred by a Uniform Resource Locator (URL) as such:

<a href="http://www.microsoft.com/about.aspx"> click me </a>

The actual URL is what is between the quotes. In its simplest form it starts with the protocol http, then the website domain name which is eventually mapped into the IP of the web server and a file path that specify which web page to display from the website www.microsoft.com

### 7.1. Web Crawler

Web crawling is the process by which we gather pages from the Web to index them and support a search engine. The objective of crawling is to quickly and efficiently gather as many useful web pages as possible, together with the link structure that interconnects them [6]. A *crawler* which is sometimes referred to *spider*, *bot*, or *agent* is a software whose purpose it is perform web crawling. Figure 4 illustrates the job of a web crawler and its role in building search engines.

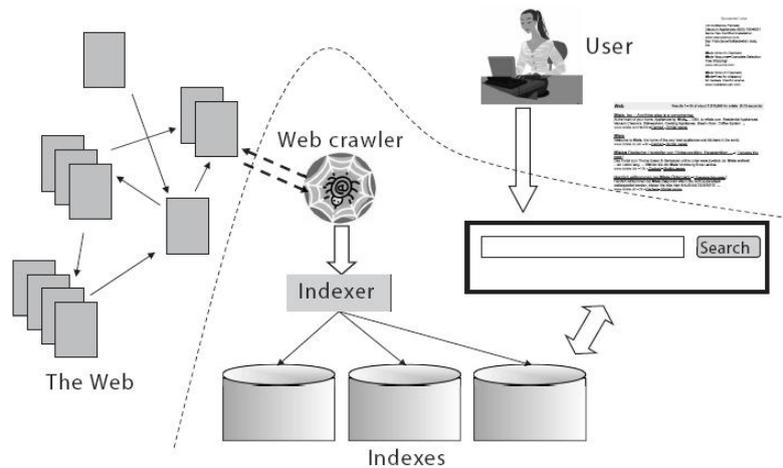

Figure 4 – Web Crawling

The architecture of a crawler as a single component is in fact composed of five modules which are depicted in Figure 5.

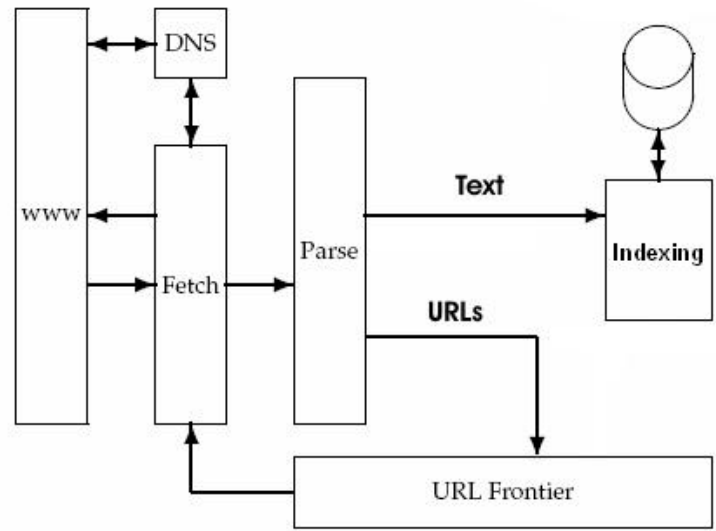

Figure 5 – Crawler Architecture

1. The URL frontier, containing URLs yet to be fetched in the current crawl
2. A DNS resolution module that determines the web server from which to fetch the page specified by a URL.
3. A fetch module that uses the http protocol to retrieve the web page at a URL.
4. A parsing module that extracts the text and the hyperlinks from a fetched web page.
5. An Indexer that stored the retrieved text along with the corresponding URL of the website being indexed in a database.

At first, the web crawler starts from a predefined set of URLs called *seed set* which should be initially stored in the *URL frontier*. The *fetch* module reads a particular URL from the frontier and uses the HTTP protocol to fetch it from the World Wide Web. It resolves its IP with the help of the DNS module. The fetch module then returns the web page in hypertext format which is then parsed by the parser module to extract the content of the webpage and the hyperlinks contained within. The content, normally a text, is indexed in a database along with the URL of the webpage. The extracted hyperlinks are stored in the URL frontier so as to the cycle starts all over again.

### 7.2. Special Topics in Web Crawling

Many hosts on the web restrict some of their pages from being crawled and therefore they are excluded by the web crawler. This standard is known as the *robots exclusion protocol*. This is done by placing a file with the name robots.txt at the root of the URL hierarchy at the site. *Robots.txt* contains hand written instructions for the crawler to exclude some particular pages from being crawled.

Another issue is that while being fetched, URLs should be *normalized* before they get parsed. Often URLs in the HTML are relative links and not absolute. Hence URLs must be converted first to an absolute link form. For instance, let's say the webpage being crawled is www.microsoft.com/index.html. Any occurrence in index.html of *<a href="about.html">About us</a>* must be converted to *<a href="http://www.microsoft.com/about.html" >About us</a>*.

Finally, the URL should be checked for *duplicate elimination*; If the URL is already in the frontier or it was already crawled, it should not be added to the frontier.

## Acknowledgment

This research was funded by the Lebanese Association for Computational Sciences (LACSC), Beirut, Lebanon, under the "Web Information Retrieval Research Project – WIRRP2011".

## References

[1] Doyle Lauren, Joseph Becker, "Information Retrieval and Processing", Melville, 1975.

[2] Singhal, Amit, "Modern Information Retrieval: A Brief Overview", Bulletin of the IEEE Computer Society Technical Committee on Data Engineering, 2001.

[3] Jurafsky and Martin, "Speech and Language Processing", Prentice Hall, 1999.

[4] Gerard Salton, A. Wong, C. S. Yang, "A Vector Space Model for Automatic Indexing", CACM 18(11), 1975.

[5] Luhn, H.P., "A Statistical Approach to Mechanized Encoding and Searching of Literary Information", IBM J. Res. Develop, 1957.

[6] Manning, Raghavan, Schutze, "Introduction to Information Retrieval", Cambridge University, 2008

[7] Feldman, Sanger, "The Text Mining Handbook - Advanced Approaches in Analyzing Unstructured Data", Cambridge University, 2007.

[8] Bing Liu, "Web Data Mining - Exploring Hyperlinks, Contents, and Usage Data", Springer-Verlag Berlin Heidelberg, 2007.

[9] Y. Yang, and X. Liu. "A Re-Examination of Text Categorization Methods", In Proc. of the ACM SIGIR Intl. Conf. Research and Development in Information Retrieval (SIGIR'99), pp. 42–49, 1999.